\documentclass[seceq]{ptptex}

\usepackage[]{graphicx}

\newcommand{\half}{\frac{1}{2}}
\newcommand{\bright}{\begin{flushright}}
\newcommand{\eright}{\end{flushright}}
\newcommand{\bminip}{\begin{minipage}}
\newcommand{\eminip}{\end{minipage}}
\newcommand{\bcent}{\begin{center}}
\newcommand{\ecent}{\end{center}}
\newcommand{\beq}{\begin{equation}}
\newcommand{\eeq}{\end{equation}}
\newcommand{\beqa}{\begin{eqnarray}}
\newcommand{\eeqa}{\end{eqnarray}}
\newcommand{\barr}{\begin{array}}
\newcommand{\earr}{\end{array}}
\newcommand{\bfig}{\begin{figure}}
\newcommand{\efig}{\end{figure}}
\newcommand{\bpc}{\begin{picture}}
\newcommand{\epc}{\end{picture}}

\renewcommand{\theequation}{\arabic{section}.\arabic{equation}}
\newcommand{\nnb}{\nonumber}
\newcommand{\reflef}{(\ref}
\newcommand{\MP}{M_{\rm P}}

\newcommand{\Lmd}{\Lambda}
\newcommand{\bra}{<\hspace{-.2em}}
\newcommand{\ket}{\hspace{-.2em}>}
\newcommand{\dbra}{<\hspace{-.45em}<}
\newcommand{\dket}{>\hspace{-.42em}>}

\newcommand{\lsim}{\mbox{\raisebox{-.3em}{$\;\stackrel{<}{\sim}\;$}}}

\title{
An approach toward the laboratory search
for the scalar field as a candidate of Dark Energy
}

\author{
Yasunori \textsc{Fujii}$^{a}$ and Kensuke \textsc{Homma}$^{b,c}$
}

\inst{
$^{a}$ Advanced Research Institute for Science and Engineering, Waseda
University,\\
Okubo, Tokyo, 169-8555 Japan\\
$^{b}$ Graduate School of Science, Hiroshima University, Japan, \\
Higashi-Hiroshima 739-8526, Japan\\
$^{c}$ Ludwig-{\cal M}aximilians-Universit\"{a}t {\cal M}\"{u}nchen, Fakult\"{a}t
f. Physik, Am Coulombwall\\
1, D-85748, Germany
}

\abst{
The observed accelerating universe indicates the presence of Dark Energy
which is probably interpreted in terms of an extremely light gravitational
scalar field.  We suggest a way to probe this scalar field which
contributes to optical light-by-light scattering through the resonance
in the quasi-parallel collision geometry.  As we find,  the frequency-shifted
photons with the specifically chosen polarization state can
be a distinct signature of the scalar-field-exchange process in spite of the
extremely narrow width due to the gravitationally weak coupling to photons. 
Main emphasis will be placed in formulating a prototype theoretical
approach, then showing how the weak signals from the gravitational
coupling are enhanced by other non-gravitational effects at work in
laser experiments.   
}

\begin{document}

\maketitle

\section{Introduction}

The discovery of the accelerating universe  \cite{exp} left today's
version of the 
cosmological constant problem, mainly consisting of a pair of
questions; the fine-tuning  problem and the coincidence problem.  Most
promising to understand them appears to introduce a scalar field
\cite{quint,cup,AT}, particularly in accordance with the way of Jordan's
scalar-tensor theory (STT) \cite{jordan}, one of the best-known
alternatives to Einstein's General Relativity.  
We add a cosmological constant $\Lmd$  as a new
ingredient, technically in the so-called Jordan (conformal) frame.
Unlike in any other approaches, we then derive the scenario of a
decaying cosmological constant in the Einstein frame corresponding to
the observation;\footnote{This relation implies an {\em overall}
behavior of the effective $\Lmd$, superimposed on which we expect {\em
local} plateaus occurring sporadically, hence mimicking {\em constants}
during certain duration of time, as illustrated in  Fig. 5.8 of
\cite{cup}.} $\Lmd_{\rm obs}\sim t^{-2}$, where the present age of the
universe $t_0 =1.37 \times 10^{10}{\rm y}$ is re-expressed as $\sim
10^{60}$ in the reduced Planckian units with $c=\hbar =\MP (=(8\pi
G)^{-1/2}\sim 10^{27}{\rm eV})=1$.   Given the unification-oriented
expectation   $\Lambda \sim 1$ in these units, the decaying behavior
provides us with the way of understanding naturally why the observed value is 
as small as $10^{-120}$.  The resulting number is this small only because
we are {\em old} cosmologically {\em not} because we fine-tune any of
the theoretical parameters.   For more details see the Refs
\cite{cup,yfob,yfptp,ipmurio,yfgrg}.

The scalar field, denoted by $\phi$,\footnote{We are going to use the
symbol $\phi$ for the canonical scalar field in the Einstein frame, rather than
in the Jordan frame; different ways in \cite{cup} in which $\phi$ and
$\sigma$ represent the scalar fields in the Jordan and the Einstein
frames, respectively. } in STT is then expected to fill
up nearly 3/4 of the entire cosmological energy \cite{exp}, known as
Dark Energy (DE).  In addition to this aspect in which the scalar field
plays a major role in the evolution of the entire universe, it also
mediates a force between local objects.  Unlike the former, the latter
component behaves in accordance with the relativistic quantum field theory
on the local tangential Minkowski spacetime, now suggesting an experimental
way to search for it.

It  couples with other microscopic fields basically as weakly as
gravity. It also shows  no immunity against acquiring a nonzero mass  due
to the self-energy, unlike genuine gauge fields like photon and
graviton.  A simple one-loop diagram in which the light quarks and
leptons with a typical mass $m_{\rm q}\sim {\rm MeV}$ couple to the
scalar field with the gravitational coupling with the strength $\sim
\MP^{-1}$ produces the mass $m_\phi$ given by
\beq
m_\phi^2 \sim \frac{m_{\rm q}^2 M_{\rm ssb}^2}{M_{\rm P}^2}\sim (10^{-9}{\rm eV})^2, 
\label{mass1}
\eeq
where we have included the effective cutoff coming from the
super-symmetry-breaking 
mass-scale $ M_{\rm ssb} \sim {\rm TeV}$,
though allowing a latitude of  the few orders of magnitude.  The
force-range turns out to be $m_\phi^{-1}$ which has a macroscopic size
$\sim 100{\rm m}$ \cite{cup,nat}.

Past searches for the scalar force of this kind have been plagued by its
matter coupling basically as weak as gravity, hence calling often for
heavy and huge objects, sometimes even natural environments, including 
reservoirs, bore holes, polar ice and so on, with so many uncontrollable 
 uncertainties \cite{wepv}.  This blockade can be avoided, however,  by
 appreciating that certain scattering amplitude in which $\phi$ occurs
 as a resonance reaches a kinematical maximum independent of the
 interaction strength.  The question is, however, what scattering system
 accommodates $\phi$ as light as above.   We may focus upon the
 2-photon system as the  most convenient candidate.  We also point out
 that STT allows $\phi$ to couple to the photons  only at the cost of
 violating the weak equivalence principle (WEP) \cite{yfms}, though
 without offending the core of spacetime geometry in General
 Relativity.\footnote{The reader is advised to consult Chapter 1.3.1 of
 \cite{cup}.}  We must confront, at the same time, the resonance width
 which should be very narrow if the coupling is gravitationally weak.

Through detailed study of the two-photon dynamics in which the
photon-photon scattering amplitude is dominated by the
$\phi$-resonance to a very good approximation, to be referred to as
$\phi$-{\em resonance-dominance},  
we are going to outline how we can enhance the gravitationally weak
signals by non-gravitational effects, described  in terms of a few
number of steps each of which is highly nontrivial, including the recent
achievements of laser technology.

We also point out that our approach is somewhat similar to the axion
search \cite{axion} in which the pseudoscalar field is produced in the
real state rather than in the virtual state.  In the present article, however,
we attempt detailed discussion on wider aspects of the scattering
system, including the physical effects of a narrow resonance, angular
distribution, particularly the behaviors in the extremely forward
direction .

  
\section{Kinematics}
\setcounter{equation}{0}

\begin{figure}
\bcent
\includegraphics[width=7.0cm]{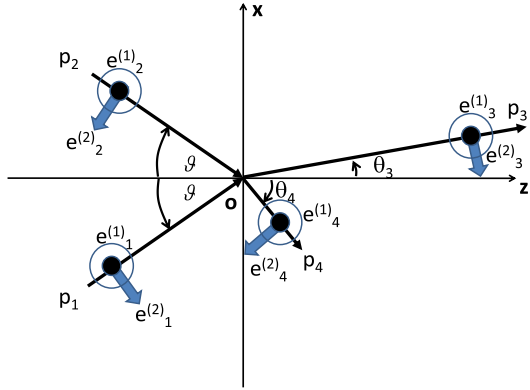}
\caption{
Definitions of kinematical variables.
}
\label{Fig1}
\ecent
\end{figure}

For the reasons to be explained shortly, we prefer a special
coordinate frame, as shown in Fig.\ref{Fig1}, in which two 
photons labeled by  1 and 2 sharing the same frequency are 
incident nearly parallel to each other, making a small angle $\vartheta$ with
a common central line along the $z$ axis.  In this {\em quasi-parallel
frame}, we define the $zx$ plane 
formed by $\vec{p}_1$ and $\vec{p}_2$, with the components of the
4-momenta $p_1 =(\omega\sin\vartheta,0,\omega\cos\vartheta ; \omega)$ and $p_2
=(-\omega\sin\vartheta,0,\omega\cos\vartheta ; \omega)$.

The outgoing photons are assumed to be in the same $zx$ plane, 
to be convenient particularly in the $s$-channel reaction, showing
an axial symmetry with respect to the $z$ axis;  $p_3 =(\omega_3
\sin\theta_3, 0, \omega_3 \cos\theta_3 ; \omega_3)$, $p_4 =(-\omega_4
\sin\theta_4, 0, \omega_4 \cos\theta_4 ; \omega_4)$.   The angles $\theta_3$
and $\theta_4$, both positive $<\pi$, are defined also in
Fig.\ref{Fig1}.   This coordinate frame is reached by transforming the
conventional center-of-mass frame for the head-on collision in the $x$
direction  by a Lorentz transformation with the velocity  in
the $z$ direction of the magnitude $c  \cos\vartheta$ with the velocity
of light $c$.

The conservation laws are
\beqa
0 \mbox{-axis}:&&\omega_3 +\omega_4 = 2\omega, \label{kinm_3}\\
z\mbox{-axis}:&&\omega_3 \cos\theta_3 + \omega_4 \cos\theta_4= 
2\omega \cos\vartheta, \label{kinm_4}\\
x\mbox{-axis}:&&\omega_3\sin\theta_3 =\omega_4\sin\theta_4. 
\label{kinm_5}
\eeqa
For a convenient ordering $0<\omega_4 < \omega_3 <2\omega$, we may choose
$0<\theta_3<\vartheta<\theta_4<\pi$, without loss of generality.   From
\reflef{kinm_3})-\reflef{kinm_5}) we derive 
\beq
\frac{\sin\theta_3}{ \sin\theta_4}=\frac{\sin^2\vartheta}{ 1-2\cos\vartheta\cos\theta_4+\cos^2\vartheta}. 
\label{kinm_6}
\eeq

The differential elastic scattering cross section favoring the
higher photon energy $\omega_3$ is given by 
\beq 
\frac{d\sigma}{d\Omega_3}=\left(\frac{1}{8\pi \omega
}\right)^{2}\sin^{-4}\vartheta \left(
\frac{\omega_3}{2\omega} \right)^2 |{\cal M}|^2,
\label{kinm_13}
\eeq
where ${\cal M}$ is the invariant amplitude, and
\beq
\omega_3=\frac{\omega\sin^2\vartheta}{1-\cos\vartheta \cos\theta_3},
\label{kinm_14}
\eeq 
which shows $\omega_3$ reaching up to $2\omega$ as $\theta_3 \rightarrow
0$, as shown in Fig. \ref{Fig2}.  In other words, we have a sharp peak of the frequency-doubled final photon
with $\omega_3 \approx 2\omega$, the total energy, in the extremely forward direction
concentrated in $\theta_3 \lsim \vartheta$, or the half-width
$\vartheta$.  This is certainly a unique observational signature.

\bcent
\bfig[h]
\hspace{12.5em}
\includegraphics[keepaspectratio,width=5.0cm]{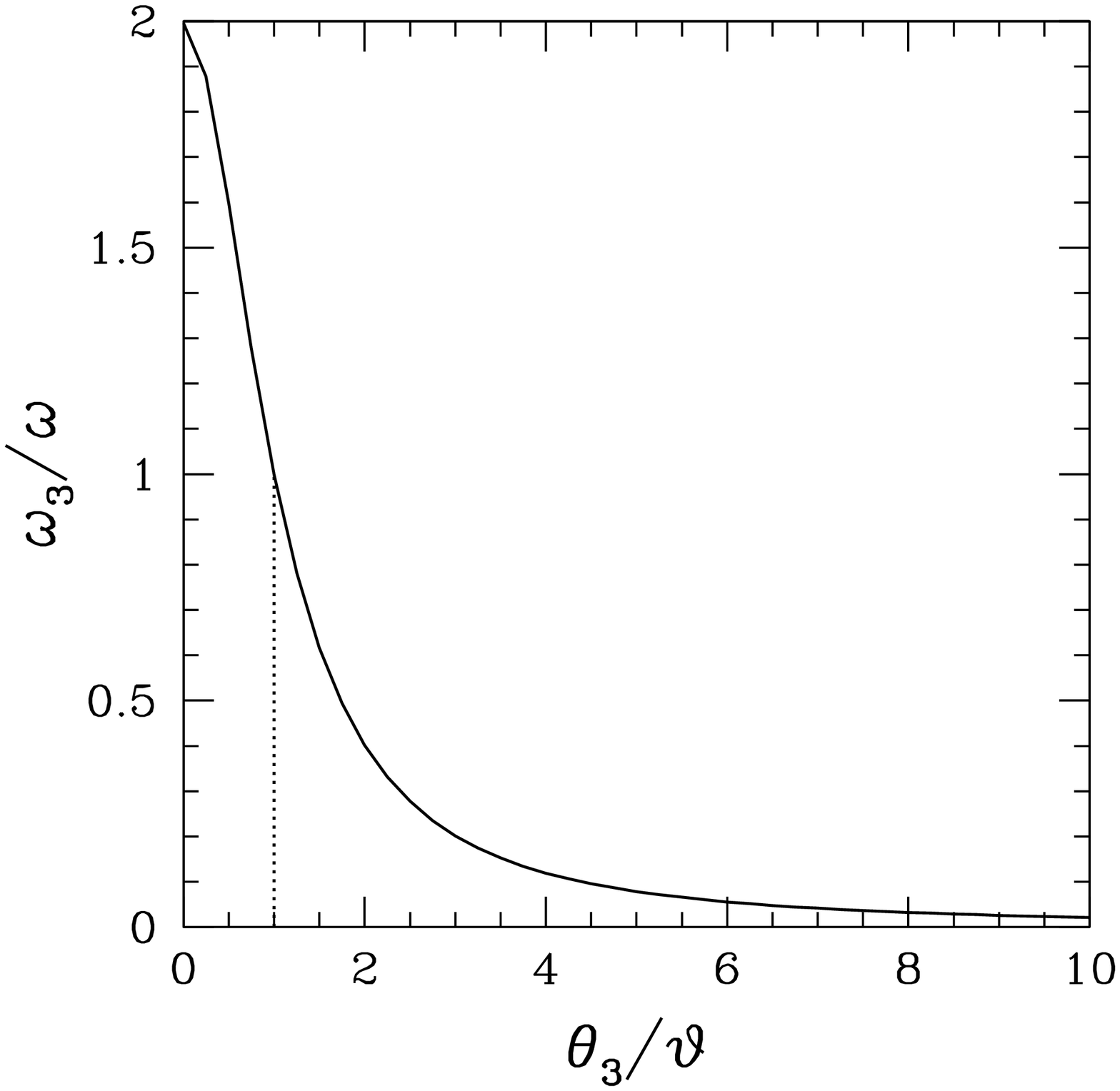}
\caption{$\omega_3 /\omega$ plotted against $\theta_3/\vartheta$.  Note
that the forward peak is extremely narrow with the angular width $\sim 
\vartheta \sim 10^{-9}$.  }
\label{Fig2}
\efig
\ecent

We point out that the factor $\sin^{-2}\vartheta$ in \reflef{kinm_13})
is derived in computing the phase-volume of the two photons in
the final state, as explained in Appendix A, while another same factor
comes from the inverse of the flux $\sqrt{(p_1p_2)^2}\approx
2\omega^2\sin^2\vartheta$ in the photon-photon scattering state in the
quasi-parallel frame, as will be discussed in Appendix E.  
As we will later discuss, the angle $\vartheta$ is going to be given by
$\vartheta \sim m_\phi /\omega \sim 10^{-9}$ for $\omega \sim
\omega_1(=1{\rm eV}$ chosen conveniently for the typical optical laser
frequency).\footnote{The symbol $\omega_1$ will be used repeatedly as a
frequency (or the energy in the units with $c=\hbar =1$) of this {\em
value}.  }   Out of the huge number 
 $\sin^{-4} \vartheta \sim \vartheta^{-4}$ as large as $\sim 10^{36}$, a
 ``half'' $\sim 10^{18}$ does indeed compensate the  gravitationally small
 number $\sqrt{m_\phi /\MP}\sim 10^{-18}$.  Unfortunately, however,
 another ``half'' of this trove will be lost because the final yield is
 reduced by  the ratio of $\vartheta^2$ when we try to measure the
 small amount of  outgoing photons of nearly doubled frequency in the
 extremely forward direction.  We  still maintain a non-trivial gain in
 the efforts  for overcoming the  gravitationally weak signals.

In this connection we find it important to point out that
there is a decisive difference between the  quasi-parallel incident direction
corresponding to $\vartheta \sim 10^{-9}$ and the truly parallel
incident beams with $\vartheta =0$.  In the lowest-order QED calculation
\cite{DG}, the gauge invariance of the theory results in the  cross
section which vanishes in the limit $\vartheta \rightarrow 0$.  The same
results are also shared   in our scalar-field dynamics, as will be
explained later toward the end of section 3.  In this context, the
negative-power dependence on $\vartheta$ as in \reflef{kinm_13}), 
for example, can be still useful for finite $\vartheta$ 
in the transitional range before reaching the truly parallel limit.


\section{Dynamics}
\setcounter{equation}{0}

\bcent
\bfig[h]
\hspace{4.0em}
\includegraphics[keepaspectratio,width=12cm]{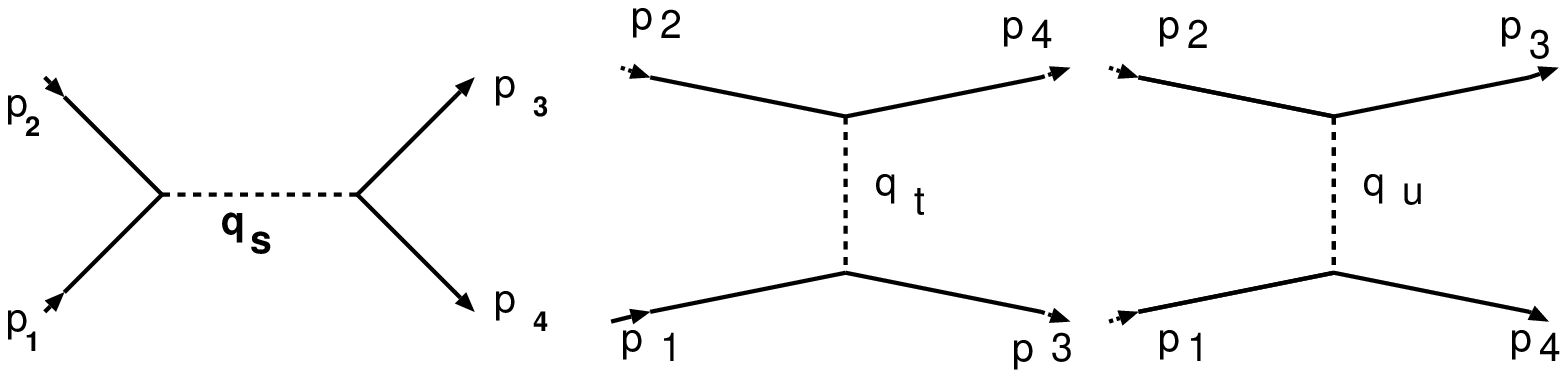}
\caption{$\phi$-dominated diagrams for the  photon-photon scattering. Solid lines are for the photons with the attached momenta $p$'s
while the dashed lines for $\phi$, in the $s$-, $t$-, and
$u$-channels, respectively. }
\label{treed}
\efig
\ecent

We assume the tree diagrams illustrated in Fig. \ref{treed}, with the
vertices given by the $\phi$-photon coupling described by the
effective interaction Lagrangian;
\beq
-L_{\rm mx\phi}=\frac{1}{4}B\MP^{-1}F_{\mu\nu}F^{\mu\nu}\phi,
\label{mxelm_1}
\eeq
where, due to the quantum-anomaly-type estimate,
$B=(2\alpha/3\pi){\cal Z}\zeta$ \cite{cup}.\hspace{-.6em}
\footnote{The  coefficient (1/12) in (6.181) in \cite{cup} has been
multiplied by 4 when the complex scalar matter fields in the loop in the toy
model are replaced by the more realistic Dirac fields.  $\zeta$ is a
constant of the order unity, while ${\cal Z} =5$  is the effective
number of the fundamental charged particles in the loop, quarks and
leptons, actually the sum of their squared charges but re-normalized in
units of the standard electron charge squared.  }
Note that the  coupling constant $\sim \MP^{-1}\sim G^{1/2}$ implies
being {\em weak} gravitationally.  The above interaction term, which has
been discussed also from a phenomenological point of view \cite{bek}, is
WEP violating \cite{yfms},\footnote{This failure of WEP is a typical
quantum effect, thus causing the true physical frame being slightly off the
pure Einstein frame, hence producing two kinds of consequences: (i)
non-Newtonian-type interactions, discussed in Chapter 6.4 of \cite{cup};
(ii) the electron mass which provides with the Bohr radius and the
atomic-clock frequency is maintained constant in the Einstein frame to a
good approximation, 
in practice, to observe Own-unit-insensitivity principle \cite{yfptp}.  }
already against Brans-Dicke's premise \cite{bd}.

We find, for example, 
\beq
<0\:| F_{\mu\nu}|\:p_1,e_1^{(\beta)}\!>=i\left(p_{1\mu}e_{1\nu}^{(\beta)} -
p_{1\nu}e_{1\mu}^{(\beta)}\right),
\label{mxelm_4}
\eeq
giving the two-photon decay rate of $\phi$ with the mass $m_\phi$;
\beq
\Gamma_\phi
=(16\pi)^{-1} \left(
B\MP^{-1}\right)^2 m_\phi^3,
\label{mxelm_4a}
\eeq
by assuming purely elastic scattering.  We find  the lifetime
$\tau_\phi =\Gamma_\phi^{-1}$  to be as long as $\sim 3\times
10^{54}$ times the present age of the universe.

The polarization vectors  are given by $\vec{e}_i^{\
(\beta)}$ with $i=1, \cdots,4$ for the photon labels, whereas
$\beta =1,2$ are for the kind of the linear polarization, also shown in
Fig.\ref{Fig1}.

In the $s$-channel, as illustrated in the first diagram in
Fig. \ref{treed}, the scalar field is exchanged between the pairs
$(p_1, p_2)$ and $(p_3, p_4)$, thus giving the squared momentum
of the scalar field $q_s^2 =\left(p_1+p_2\right)^2 =2\omega^2 \left(
\cos 2\vartheta -1 \right) <0$ with the metric convention $(+++-)$.

With the type $\beta=1$ for all the photons we find;\footnote{The four
digits in the subscript are for $\beta$ arranged from left to right
according to the photon labels, 1-4.}
\beq
{\cal M}_{1111s}=-(B\MP^{-1})^2\frac{\omega^4 \left(
\cos2\vartheta -1\right)^2}{2\omega^2 \left(
\cos2\vartheta -1\right)+m_\phi^2},
\label{mxelm_7}
\eeq
also with ${\cal M}_{1111s}={\cal M}_{2222s}=-{\cal M}_{1122s}=- {\cal
M}_{2211s}$ for the only nonzero components.

It immediately follows that the denominator, the propagator of $\phi$, vanishes at
\beq
\omega^2 =\omega_r^2 \equiv  \frac{m_\phi^2/2}{1-\cos 2\vartheta},
\label{mxelm_7_1}
\eeq
corresponding to the creation of the scalar field $\phi$.  Note that
with the assumed value of $m_\phi \sim 10^{-9}{\rm eV}$, we find that
$\omega_r$ can be as large as $\sim \omega_1 =1{\rm eV}$ by choosing
$\vartheta \sim 10^{-9}$, as alluded before.\footnote{This relation may
also be interpreted that the quasi-parallel frame provides us with a
device that {\em lowers} the center-of-mass energy to the invariant mass
$m_\phi$ by choosing $\vartheta$ small enough for given $\omega$. }

 At the same time the decay rate \reflef{mxelm_4a}) suggests that the
 propagator acquires an imaginary part obtained approximately by the
 replacement; $m_\phi^2 \rightarrow \left( m_\phi -i\Gamma_\phi/2
 \right)^2 \approx m^2_\phi -im_\phi \Gamma_\phi $.  We may then find
 that the denominator, denoted by ${\cal D}$, can be re-expressed as 
\beq
{\cal D} \approx 2\left( 1-\cos2\vartheta  \right)\left( \chi+ia
\right),\quad\mbox{with}\quad \chi =\omega^2 -\omega_r^2,
\label{mxelm_7_2a}
\eeq
where
\beq
a=\frac{m_\phi \Gamma_\phi/2}{1-\cos 2\vartheta}.
\label{mxelm_7d}
\eeq
Choosing $\vartheta \sim 10^{-9}$, also combining again with
\reflef{mxelm_4a}), we find $a \sim 10^{-77}({\rm eV})^2$, which will be
used later repeatedly.

The amplitude \reflef{mxelm_7}) can be further re-expressed by
the Breit-Wigner formula.  To show this explicitly, we denote the
portion of \reflef{mxelm_7}) other than ${\cal D}$ by ${\cal N}$
evaluated at $\omega_r$;
\beq
{\cal N}\approx \left( B\MP^{-1} \right)^2 \omega_r^4 \left( 1-\cos
2\vartheta \right)^2.  
\label{mxelm_7_2}
\eeq
We compare this with \reflef{mxelm_4a}), eliminating the common factor $\left(
B\MP^{-1} \right)^2$, to obtain
\beq
{\cal N}\approx 16\pi \Gamma_\phi m_\phi^{-3}\omega_r^4\left( 1-\cos
2\vartheta \right)^2 =4\pi \Gamma_\phi m_\phi=8\pi a\left( 1-\cos
2\vartheta \right),
\label{mxelm_7_3}
\eeq
where we have used \reflef{mxelm_7_1}) and \reflef{mxelm_7d}) in the
second and the last steps, respectively.  From this and
\reflef{mxelm_7_2a}), we finally obtain
\beq
{\cal M}_{1111s}=\frac{\cal N}{\cal D}\approx  4\pi\frac{a}{\chi+ia},\quad\mbox{hence}\quad 
|{\cal M}_{1111s}|^2 \approx  (4\pi)^2 \frac{a^2}{\chi^2+a^2}.
\label{mxelm_13}
\eeq

At the resonance position $\omega=\omega_r$ or $\chi=0$, we find ${\cal
M}_{1111s}=-4\pi i$ or $|{\cal M}_{1111s}|^2=(4\pi)^2$, 
a ``large''value entirely free from being small for a gravitational force.
This obvious result due directly to the element of quantum mechanics is
the heart of what we may call $\phi$-dominance, as was
pointed out in section 1.   In fact without this resonance we would have
found $|{\cal M}|^2$ to be as small as $\MP^{-4}$ either directly from the
diagram due to \reflef{mxelm_1}) or from the numerator $a^2$ in the
second of \reflef{mxelm_13})  again combined with \reflef{mxelm_4a}).

This smallness holds true  for the processes in the $t$- and $u$-channels, as
well, with $q_t$ and $q_u$ both spacelike, hence no chance of being 
enhanced due to the resonance.  In other words, a resonance occurring
only in the $s$-channel awards us with a {\rm gain} of $\sim \MP^4$, 
or of such dimensionless factors like $(\MP/ ({\rm eV}))^4\sim10^{108} $
 or $(\MP/m_\phi)^4\sim 10^{144}$ for $m_\phi \sim 10^{-9}{\rm eV}$.
  We may take  advantage of such huge numbers of this nature in our
effort to overcome the weak coupling of gravity, as alluded at the
beginning.  We face, however, another consequence of the weak coupling
resulting in the extremely {\em narrow} width $a \sim 10^{-77}({\rm
eV})^2$, implied by $\MP^{-2}$, unimaginably small in   any practical
measurements available currently.

This is even narrower than what is expected from the quantum-theoretical  
uncertainty of the momenta to be included in any of the existing beams 
in  particle physics.
The quasi-parallel frame, as we noted before, requires us  
to prepare the beam with the accuracy of the angle 
$\vartheta \sim 10^{-9}$ to be realized around the focal point 
in the diffraction limit, where the momentum uncertainty is unavoidable
in principle.  This seems to present an argument, though never stated before, to be  applied carefully to the extreme situation under the present discussion.
A natural way to cope with the issue of this fundamental 
importance is to apply an {\em averaging} process over  
the range of the likely 
uncertainty. More specifically, we integrate the squared amplitude  
with respect to $\chi$ uniformly over the range 
${\cal R}=(-\tilde{a},\tilde{a})$;
\beq
\overline{|{\cal M}_{1111s}|^2}=\frac{1}{2\tilde{a}}\int_{-\tilde{a}}^{\tilde{a}}
|{\cal M}_{1111s}|^2 d\chi =(4\pi)^2 \eta^{-1}\frac{\pi}{2}\hat{\eta}.
\label{mxelm_14}
\eeq
The far RHS is obtained immediately by substituting from the second of
\reflef{mxelm_13}), where we assume $\eta \equiv \tilde{a}/a \gg 1$, also with
$\hat{\eta}=(2/\pi)\tan^{-1}\eta$, reaching the maximum 1 for $\eta
\rightarrow \infty$, while reducing to be negligibly small if the
resonance is outside the range ${\cal R}$.  We emphasize that the
integral in \reflef{mxelm_14}) is  insensitive to the  choice of the
integration boundaries, as long as  they are much larger than $a$.  We
may also find explicitly how the uncertainty in $\vartheta$ affects the
same in $\chi$ will be shown shortly in \reflef{exeq_2ccc}), for example.

The averaging process proposed in \reflef{mxelm_14}) has an added 
advantage, by providing a practical way of measurement of the realistic
accuracy of the order $\sim {\rm eV}$.  We also notice that the small number
$\eta^{-1}=a/\tilde{a}$ on the right-hand side of  \reflef{mxelm_14})
simply reflects   how small a portion  the  resonance occupies in the
entire    range ${\cal R}$.  Part of the  gain as large as than $10^{144}$
 emphasized above will  thus be offset by $\eta^{-1}$.  We nevertheless
 will show that the net result is still sufficiently large.  For this
 purpose, we start with summarizing what we have found so far, by
 substituting \reflef{mxelm_14}) into \reflef{kinm_13}) yielding the
 averaged cross section
\beq
\overline{\left( \frac{d\sigma}{d\Omega_3} \right)_s}
=\frac{\pi}{8\omega^2}\left( \frac{\omega_3}{2\omega}  \right)^2\vartheta^{-4} \eta^{-1},
\label{mxelm_19}
\eeq
where we have put $\hat{\eta}=1$.

We want to relate the boundary value $\tilde{a}$ somehow to the
observations.  For this purpose we first notice that in the above
analysis it may sound as if we vary the incident frequency $\omega$
 with the angle $\vartheta$ kept fixed to the value $\vartheta_1$, for
 example, so that, according to  \reflef{mxelm_7_1}),  
\beq
\omega_r^2=\frac{m_\phi^2}{4\vartheta_1^2},
\label{mxelm_19_1}
\eeq
which is hence fixed.  In the process of beam focusing with a laser
field, on the contrary, we are led almost to fix $\omega$  to
$\omega_1(=1{\rm eV}, \mbox{for example})$, but leaving $\vartheta$ to vary
 largely due to the unavoidable nature in the diffraction limit.  
Then  $\omega_r$ is now a variable expressed  as a function of $\vartheta$ by
\beq
\omega_r^2=\frac{m_\phi^2}{4\vartheta^2}.
\label{mxelm_19_3}
\eeq
Substituting this into the second of \reflef{mxelm_7_2a}), we obtain
\beq
\chi(\vartheta)= \omega_1^2 \left( 1-\left(\frac{\vartheta_r}{\vartheta}\right)^2 \right),
\label{exeq_2ccc}
\eeq
where
\beq
\omega_1^2 =\frac{m_\phi^2}{4\vartheta_r^2},
\label{mxelm_19_4}
\eeq
which defines $\vartheta_r$.\footnote{The alternate roles played by
$\omega$ and $\vartheta$ in this sense indicate a more general situation
that they represent  two experimental handles of the system.  The
resonance takes place by satisfying the relation of the type of one of
\reflef{mxelm_19_1}), \reflef{mxelm_19_3}) or \reflef{mxelm_19_4}),
which define the resonance condition expressed by a {\em band} in the
$\omega$-$\vartheta$ plane.   }

Suppose the two ends $\pm \tilde{a}$, obviously chosen for simplicity, in \reflef{mxelm_14}) correspond to
the two boundary values of the angle $\vartheta_\pm$,
\beq
\chi(\vartheta_\pm)=\pm \tilde{a},
\label{mxelm_19_5}
\eeq 
which allows us to express $\vartheta_-$ in terms of
$\vartheta_+$, as will be shown in Appendix B, eventually to give the
coefficient $\eta$ which occurs  in  \reflef{mxelm_19}); 
\beq
\eta =\left( 1- \left( \frac{\vartheta_r}{\vartheta_+}
\right)^2\right)\eta_0, 
\label{exeq_63}
\eeq
where
\beq
\eta_0 \equiv \frac{\omega_1^2}{a}\sim 10^{77}.
\label{exeq_63_1}
\eeq

We may identify $\vartheta_+$ with  the observational uncertainty
$\Delta\vartheta$  of  the angle $\vartheta$.  We may reasonably assume
$\vartheta_r \ll \Delta\vartheta$, to arrive at a simpler result
\beq
\eta \approx \eta_0, 
\label{app_b7}
\eeq
indicating $\tilde{a}$ to be of the size of practical measurements.

As far as the connection with observables are concerned, we may try
another average over $\vartheta$ rather than $\chi$ in \reflef{mxelm_14}).
As we find in Appendix C, however, the same process over $\vartheta$
results in  
\beq
\frac{\overline{|{\cal M}_{1111s}|_\vartheta^2}}{\overline{|{\cal
M}_{1111s}|^2}},
\approx \frac{\vartheta_r}{\vartheta_+},
\label{exeq_63_2}
\eeq
which implies the $\vartheta$-average somewhat smaller than that of the
$\chi$-average, still within basically of the same order of magnitude as
$\eta_0^{-1}$.

Before closing this section, 
we show explicitly that no infinity 
occurs physically as $\vartheta \rightarrow 0$, as alluded at the end of
section 2.  For this purpose, we first consider the limit in which the
$\vartheta$-dependent term $2\omega^2 (\cos 2\vartheta-1)\sim
-4\omega^2\vartheta^2$ in the denominator of \reflef{mxelm_7}) is
negligibly smaller than another term of $m_\phi^2$.  The condition may
be expressed as
\beq
\vartheta^2\ll \vartheta^2_r = \frac{m_\phi^2}{4\omega_1^2} \sim \left( 10^{-9} \right)^2,
\label{angav_5}
\eeq
where $\vartheta_r$ satisfies the resonance condition
for any choice of $\omega_1$ in the range of optical frequency $\sim 1$eV.
In this limit,
we find that the amplitude \reflef{mxelm_7}) depends on $\vartheta$ only
through the numerator, hence yielding the  behavior like $\vartheta^4$
as $\vartheta \rightarrow 0$.  It then follows that $|{\cal M}|^2$
behaves like $\sim \vartheta^8$ which overcancels $\vartheta^{-4}$ in 
\reflef{kinm_13}), hence, proving the absence of infinities
as $\vartheta\rightarrow 0$, in conformity with the lowest-order QED
calculation.  We re-emphasize, however, this occurs for $\vartheta$
much away from $\vartheta_r \sim 10^{-9}$. On the other hand,
for $\vartheta$ close to $\vartheta_r \sim 10^{-9}$ which is
$\phi${\it -resonance-dominance} not obeying \reflef{angav_5}), 
the terms like
$\vartheta^{-4}$ in \reflef{kinm_13}) can be simply large but stay finite.
This is what we may expect as one of the enhancement mechanisms.


\section{An overall enhancement scenario}
\setcounter{equation}{0}

We are now ready to sketch briefly what we have achieved based on the relation
\reflef{mxelm_19}) with $\eta\sim \eta_0 $ which applies to the simple 
models of uniform distribution with respect to $\chi$ to a reasonable
accuracy.  As we recall, we compare $|{\cal M}|^2\sim \MP^0$  at the
resonance position with  $|{\cal M}|^2 \sim \MP^{-4}$ as a whole.  This
might be represented by a gigantic leap shown near the left end of
Fig. \ref{fig:enhance}.   This will be followed by a setback 
by $\eta\sim 10^{-77}$ occurring corresponding to the averaging
processes, required to overcome the narrow width of the resonance.
Remarkably enough we are still at the middle, up by more than 70 orders
of magnitude from the bottom. 
\bfig[h]
\bpc(300,105)(-85,1.5)
\put(15,15){\line(1,0){248}}
\put(60,105){\line(1,0){60}}
\put(113,60){\line(1,0){49}}
\put(154,71){\line(1,0){60}}
\put(203,105){\line(1,0){60}}
\put(199,71){\vector(1,2){16.9}}
\put(30,15){\vector(1,2){45}}
\put(105,105){\vector(1,-2){23}}
\put(158,60){\vector(1,2){5.6}}
\put(-19,11){$\MP^{-4}$}
\put(-19,56){$\MP^{-2}$}
\put(-19,101){$\MP^{0}$}
\put(49,38){\normalsize  height of resonance}
\put(116,90){\normalsize width of resonance}
\put(218,90){\normalsize intense laser}
\put(164,60){\normalsize $\vartheta^{-2}$}
\epc
\caption{Schematic representation of enhancing the gravitationally
weak signals $\sim \MP^{-4}$ finally to those at the level of $\sim
\MP^{0}$. }
\label{fig:enhance}
\efig

We then naturally wonder if we we further exploit $\vartheta^{-4} \sim 10^{36}$
in \reflef{mxelm_19}) for a more gain. Further developing our account at the
end of the preceding section, we should scrutinize the extremely sharp
peak of the angular distribution of $\omega_3$ in the forward
direction as illustrated in Fig.\ref{Fig2}.
In this connection we are concerned about the noise due to the
large amount of non-interacting photons flowing out to the forward
direction.  One of the ways to remove this is to set up a threshold
$\bar{\omega}_3$, collecting the desired data by accepting those only with
$\omega_3 >\bar{\omega}_3$ with $\omega <\bar{\omega}_3 <2\omega$.

By substituting $\bar{\omega}_3$ into $\omega_3$ in \reflef{kinm_14}), we
 find the maximum angle $\bar{\theta}_3$ corresponding to $\bar{\omega}_3$;
\beq
\underline{x}\equiv \cos\bar{\theta}_3\approx
1+\half \vartheta^2\left( 1-\frac{2\omega}{\bar{\omega}_3}  \right)\approx
1-\frac{1}{4}\vartheta^2 u,\quad\mbox{or}\quad \bar{\theta}_3
\approx \vartheta\sqrt{\frac{u}{2}},
\label{angav_13}
\eeq
where the parameter $u$, chosen reasonably smaller than 1, is defined by
\beq
\frac{\bar{\omega}_3}{\omega}=2-u. 
\label{angav_14}
\eeq

Nearly automatically, we are now considering the {\em partially integrated} cross section defined by
\beq
\overline{\sigma}=2\pi
\int_0^{\bar{\theta}_3}\left(\frac{d\sigma}{d\Omega_3}\right)\sin\theta_3 d\theta_3.
\label{angav_15}
\eeq 
In view of the fact that $\bar{\theta}_3 \lsim \vartheta$, as shown by
the second of \reflef{angav_13}), much smaller
than any angular resolution of ordinary observations, we are making use only
of a tiny portion of the available solid angle, which we may prepare in
any conventional measurements.  Also noticing that the
integrand on RHS of \reflef{angav_15}) is likely linear with respect to
$\theta_3$, we easily expect $\overline{\sigma}$ is proportional to
$\vartheta^2$.

The exact value of the coefficient will be determined by
\beq
\int_0^{\bar{\theta}_3} \left( \frac{\bar{\omega}_3}{2\omega} \right)^2
\sin\theta_3 d\theta_3\approx \frac{1}{4}\vartheta^2 u.
\label{angav_16}
\eeq
Details of estimating LHS, as shown in Appendix D, gives RHS.

Substituting this into \reflef{angav_15}) also from \reflef{mxelm_19}) 
we finally obtain
\beq
\overline{\sigma}\approx\frac{\pi^2}{16\omega^2}\eta^{-1}\vartheta^{-2}u.
\label{angav_17}
\eeq
Unlike $\vartheta^{-4}$ in \reflef{mxelm_19}), we have now
$\vartheta^{-2} \sim 10^{18}$ as
a result of $\vartheta^{2}$ in \reflef{angav_16}).

We are still short of somewhere around 60 orders of magnitude before
reaching the goal of $\MP^0$.  
　　　　　

\setcounter{equation}{0}

\section{Enhancement by high-intensity lasers}

We now discuss how much we can further enhance the signals by making use
of  high-intensity laser fields.  In order to evaluate the
approximate order of magnitude for the required intensity, we consider
one of the simplest conceptual setups: a single Gaussian laser beam 
with the linearly polarized state 11 focused by an ideal lens.
Inside this single beam,  two photons will scatter each other in the
$s$-channel most likely in the diffraction limit at
around the focal point. A frequency-upshifted photon is emitted
nearly in the forward direction in which we may place a prism element
to directionally separate the frequency-shifted photons from 
non-interacting photons followed by
a polarization filter to ensure the final state
polarization consistent with the scalar field exchange.
As for the scalar exchange, in principle, the transitions on the
polarization states $11\rightarrow 11$ and $11\rightarrow 22$ are possible
based on (\ref{mxelm_7}).
The requirement of $11\rightarrow 22$ should improve the
signal to noise ratio against the background from the non-interacting case 
$11\rightarrow 11$ in addition to the requirement of the frequency shift.
With  many technical arrangements yet to be scrutinized, 
the simplified concept is still useful to discuss the necessary laser 
intensity for yielding a sufficient number of frequency-shifted photons.

\bcent
\bfig[h]
\hspace{10.5em}
\includegraphics[width=8.0cm]{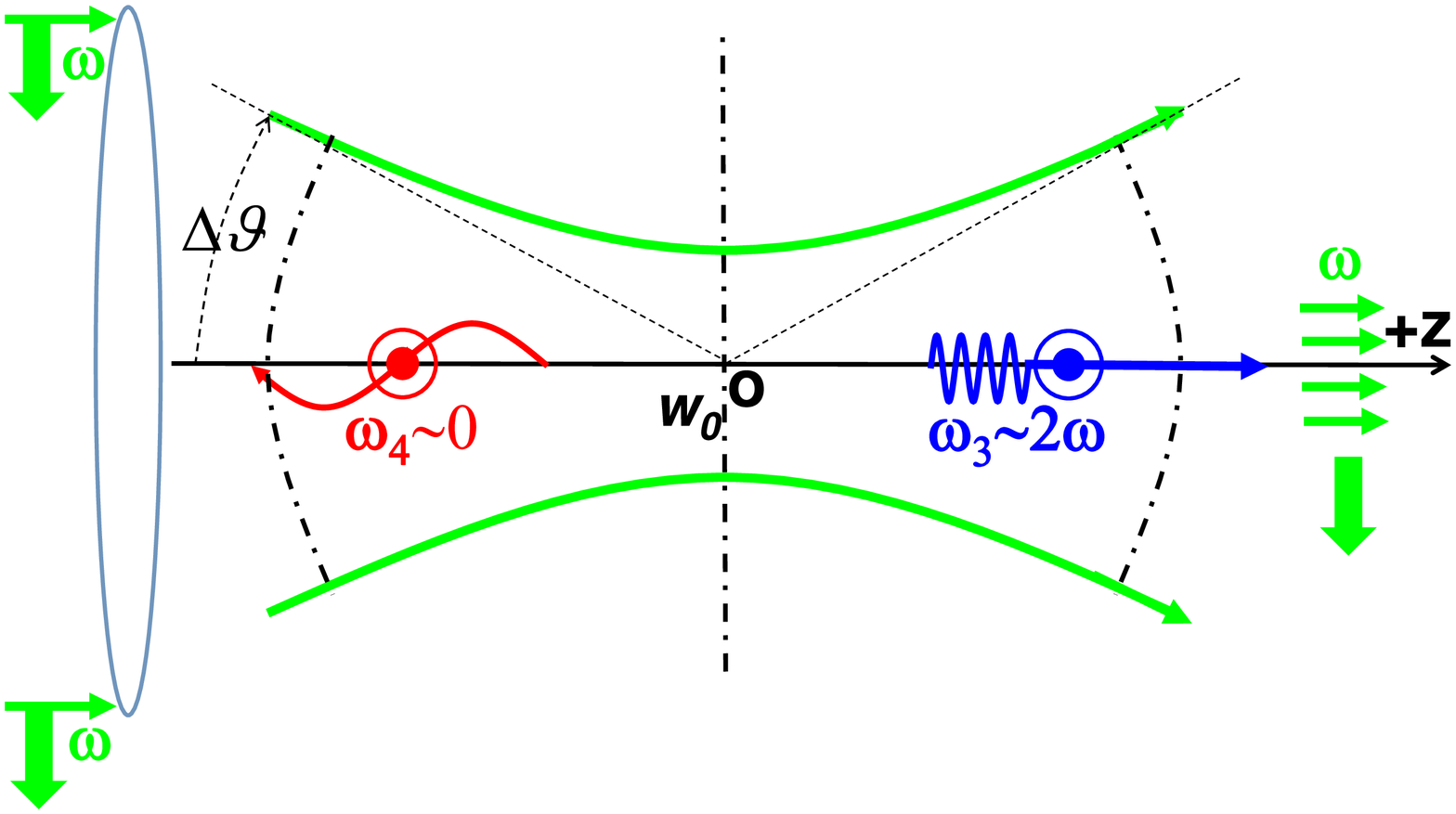}
\mbox{}\\[-2.0em]
\caption{
A single Gaussian laser beam focused by an ideal lens  
where a scalar field exchange entails a frequency-upshifted photon
in the forward direction associated with the transition of the
liner polarization state from $11$ to $22$.
The frequency of the incident laser beam is assumed to be
within a narrow band, while the incident angle varies 
largely including the value $\sim 10^{-9}$.  }
\label{onebeam}
\efig
\ecent 

Let us first consider a luminosity factor per laser pulse duration in an 
analogy to the concept applied to high-energy collider experiments.
A minimum time scale of $\sim \omega^{-1}$ is considered for interaction
within laser pulse duration.
We then define the integrated luminosity in this
time-scale only in the vicinity of the focal point by
\beq
{\cal L} = \frac{I}{\pi w^2_0},
\label{coh_1}
\eeq
where $w_0$ is the beam-waist at the focal point, while
$I=N_{\rm f}(N_{\rm f}-1)/2 \sim N_{\rm f}^2$ is the combinatorics in
choosing pairs of incident particles out of the total number $N_{\rm f}$
contained in a pulse, or a bunch.  This procedure is acceptable when we
deal with fermions which are all distinguishable from each other.  For bosons, 
like  coherent photons, however, we face a totally different
circumstance of a degenerated state in which particles are
indistinguishable, hence uncountable.  Preparing for the required new
formulation, as will be shown, turns out to entail the same intensity
factor as in the fermionic particles.\footnote{In the following, we show
a content somewhat different from section 6 of Ref. \cite{yfgrg}, though
basically as a continued outgrowth of the discussion on the inducement mechanism.  }

On the analogy with fermionic beams, we consider independent
two photon-beams which are to be collided to each other, though
more realistic experimental setup has been already proposed above.
Each of the two laser beams in this simplified case is described,
however, by means of the quantized version of the {\rm coherent state} which
features a superposition of different photon numbers, characterized by an
 {\em averaged photon number} $N$ \cite{glauber}\footnote{Instead of the
 original complex parameter $\alpha$, we use a less general real number
 $N=|\alpha|^2$ to simplify the equations shown in the following.  } 
\beq
|N\dket \equiv \exp\left
(-N/2 \right) \sum_{n=0}^{\infty}
\frac{N^{n/2}}{\sqrt{n!}} |n>,
\label{coh_2}
\eeq
where $|n>$ is the normalized state of $n$ photons
\beq
|n> =\frac{1}{\sqrt{n!}}\left( a^\dagger\right)^n |0>,
\label{coh_3a}
\eeq
with $a^\dagger$ and $a$  the creation and the annihilation
operators, respectively, of the photons assumed to share a single common
frequency and polarization. Admitting that this could be part of an 
approximation when applied to a pulse laser where multiple frequencies
must be included, we still believe it to offer a good starting
point to investigate this simplified but well-defined approach.  
We then derive immediately the normalization condition
\beq
\dbra N|N \dket=1.
\label{coh_3aa}
\eeq

From \reflef{coh_2}) we derive the relations;\footnote{These equations
were used   in \cite{glauber} to {\em define} a coherent state in the
quantum theoretical sense. } 
\beq
a|N\dket =\sqrt{N}|N\dket ,\quad\mbox{and}\quad \dbra N|a^\dagger
=\sqrt{N}\dbra N|, 
\label{coh_3c}
\eeq
where we have made use of the familiar results
\beq
 a^\dagger |n>=\sqrt{n+1}\:|n+1>,\quad\mbox{and}\quad a|n+1> =\sqrt{n+1}\:|n>.
\label{coh_3b}
\eeq
We emphasize that these relations in \reflef{coh_3b})   are the basis of
 what has been known as the {\em induced} (or stimulated)
 emission/absorption, or  creation/annihilation processes long discussed
 historically in  association with atomic
 transitions.\footnote{See \cite{IZ} for a  discussion of a
 possible enhancement effect not necessarily associated   with atomic
 transitions. }   The simplest choice $n=0$
 results in the special examples of the {\em spontaneous} processes.

By combining \reflef{coh_3c}) with \reflef{coh_3aa}) we reach
\beq
\dbra N|a| N\dket =\dbra N|a^\dagger| N\dket =\sqrt{N}.
\label{coh_3d}
\eeq
From \reflef{coh_3c}) also follows  
\beq
\dbra N|n|N\dket = 
\dbra N|\left(a^\dagger a\right)|N\dket=N,
\label{coh_3e}
\eeq
providing {\em a posteriori} derivation of the expectation value of
$n$ to be $N$.


We expect that the laser beam will enter the system in the form of a 
coherent state.  In order to understand how the coherent
state introduced above modifies the conventional Feynman amplitude, we
start with re-expressing the one for the first diagram in
Fig. \ref{treed}, with obvious simplifications;
\beq
 V_2\left( \left( p_1+p_2\right)^2+m_\phi^2 \right)^{-1}V_1,
\label{6d_1}
\eeq
where $V_1$ and $V_2$ correspond to the matrix elements of the
 interaction \reflef{mxelm_1}) at the first and the second vertices,
 respectively;
\beq
V_1 =B\bra
0|F_{\mu\nu}|p_1 \ket 
\bra 0|F^{\mu\nu} |p_2 \ket,\quad\mbox{and}\quad V_2 =B\bra
p_3|F_{\mu\nu}|0 \ket \bra p_4|F^{\mu\nu}|0 \ket.
\label{6d_2}
\eeq
For further simplicity, we suppress all the complications arising
from the momenta and polarization vectors, finding
\beq
\bra 0|F_{\mu\nu}|p_1 \ket \rightarrow \bra 0|a| p_1 \ket =1, 
\label{enhrev1_12}
\eeq
in the first factor in $V_1$,  for example, where the second of
\reflef{coh_3b}) is applied with $n=0$.

The ket $| p_1 \ket$ implies that we have a one-photon state at the
initial time.  But this feature should be revised if the photon comes from those
embedded in the coherent state prepared before the photons enter the
system from the left of the lens, as shown in Fig. 5.  The initial state
should now be $| p_1, N\dket$,  where we exhibit $p_1$ explicitly for
the momenta inside the coherent state, which is also supposed to extend
over the region around the first vertex.  The photon $p_1$ reaches here,
finding itself surrounded by the sea of photons sharing the same common
frequency and polarization as its own, thus annihilates in the induced
manner.  Correspondingly, the ``final'' state represented by $\bra 0\:|$
in \reflef{enhrev1_12}), the vacuum, is now going to be replaced by
$\dbra p_1, N|$.  In this way, the second half of \reflef{enhrev1_12})
is replaced finally by  
\beq
\dbra p_1, N|a |p_1,N\dket = \sqrt{N},
\label{enhrev1_3}
\eeq
precisely the first of \reflef{coh_3d}).  By comparing this with 
\reflef{enhrev1_12}) we come to finding that the presence of the
coherent state results in an {\em enhancement} by $\sqrt{N}$.  Basically the
same analysis applies to another incident photon with the momentum $p_2$.
 Summarizing, the presence of the coherent states enhances the matrix
 element $V_1$   by $(\sqrt{N})^2 =N$.  The same result can also be
 re-interpreted in terms of   the enhanced coupling   constant $NB$ at
 the first vertex.

We now move on to discuss $V_2$ at the second vertex, in which   the
outgoing photons are emitted with the momenta $p_3$ and $p_4$ most
likely quite different from the initial ones, in fact providing us with
what we called a unique observational signature as was emphasized  in
sections 2 and 4.  In this sense, the final photons are created
spontaneously simply from the vacuum, no enhancement, without the sea of
photons provided by the coherent state.  Combining this with the
foregoing result for the first vertex, we find the total event {\em rate}
for the nearly-doubled frequency $\omega_3$  to
be enhanced effectively by $((\sqrt{N})^2)^2=N^2$.

This result can be re-expressed as in \reflef{coh_1}), in the form of
\beq
{\cal Y}=\frac{I_{\rm c}}{\pi w_0^2}\bar{\sigma},
\label{coh_21}
\eeq
where
\beq
I_{\rm c}=1 \times N^2,
\label{coh_22}
\eeq
which, according faithfully to our derivation, should be understood to
be a product of the number (=1)  of collision between a pair of coherent states
and the coupling strength squared ($=N^2$).  It still seems intriguing
to find an approximate numerical agreement between $I_{\rm c}$ defined
by \reflef{coh_22}) and $I\sim N_{\rm f}^2$ in \reflef{coh_1}) as long
as $N \sim N_{\rm f}$, 
though they are different from each other conceptually, combinatorics {\em vs}
enhancements due to inducement.\footnote{The agreement, though 
approximate and accidental, might instigate an attempt for 
using combinatorics also applied to the coherent state in terms of the
average $N$.  This might be convenient if we insist that the
annihilated photon must have been produced before, as can be
implemented by another factor, the second of \reflef{coh_3d}).  But this 
not only lacks the responsible interaction, but is also redundant; no need for 
an ancestor.  More important is the absence of the evidence for an
overall enhancement $N^4$ in any of the laser experiments on atoms. }
See Appendix E for a brief account of the mechanism behind deriving
\reflef{coh_21}).

Now the uncertainty on the incident angle between two light waves in the
single-beam focusing is expected to be
\beq\label{exeq_3}
\Delta\vartheta \sim
\frac{w_0}{z_R} = \pi^{-1}\frac{\lambda}{w_0},
\eeq
from the definition of the Rayleigh length $z_R \equiv \pi w^2_0/\lambda$
with the 
optical laser wavelength $\lambda \sim 10^{-6}$~m~\cite{Yariv}.
In principle, we may control $\Delta\vartheta$ by changing the lens
diameter and the focal length.  For the assumed diffraction limit $w_0
\sim \lambda$, we expect $\Delta\vartheta \sim \pi^{-1}$, hence
$\vartheta_r/\vartheta_+ \ll 1$.  In this way,  we may
assume $\eta \sim \eta_0$ in \reflef{angav_17}) through the result  
\reflef{exeq_63})
for $\vartheta_r \ll \Delta\vartheta$.  
We then express the number of the expected nearly frequency-doubled
photons per pulse focusing ${\cal Y}$ as
\beq
{\cal Y}
\sim \frac{u}{64\pi}\vartheta_r^{-2}\eta^{-1} N^2.
\label{exeq_6}
\eeq
By requiring ${\cal Y}\sim 1$ per pulse focusing, 
we find the required pulse energy as
\beq
\bar{N} \sim \left( \frac{64\pi\eta\vartheta^2_r{\cal Y}}{u}\right)^{1/2}
\sim 10^{31} \mbox{optical photons} \sim 10^{10}{\rm kJ},\\[1.0em] 
\label{exeq_9}
\eeq
for $\omega = 2\pi/\lambda \sim 1 {\rm eV}$, $u \sim 0.1$, 
$\eta \sim \eta_0 \sim 10^{77}$, and $\vartheta_r \sim 10^{-9}$.
This energy per pulse is far beyond what is presently achievable.

As a rescue, we now wonder if we are allowed to revise the spontaneous 
nature at the second vertex in which both of the final photons were supposed 
to be created from the vacuum.\footnote{
Sometimes the process at
the second vertex is  called a decaying process (from the scalar field).
This might also entail a view that the two photons are the decay
products.  This may not be entirely inappropriate  from a
phenomenological point of  view.  We nevertheless continue to interpret
these photons as being created from a field-theoretical  point of view.
They are created no matter what their origin might be.  In this way we
maintain a logical consistency with understanding the process taking
place at the first vertex in the first of Fig. \ref{treed}.
}
We in fact find the desired induced nature realized 
if we introduce another laser beam again as 
a coherent state, with its momentum fine-tuned to $p_4$, which we may keep 
fixed within a range specifically in such a way to maintain the nearly-doubled
frequency $\omega_3$ in the forward direction, simply due to the
energy-momentum conservation laws. 
We may consider an experimental setup as illustrated in Fig.\ref{Fig6},
where $\omega_4$ with the prescribed energy $u\omega$
enforces $\omega_3$ to be $(2-u)\omega$.
If $\omega_3$ and $\omega_4$ are separated largely from each other,
we can define $\omega_3$ as a clear experimental signature 
by adding the specification of the polarization state consistent 
with the scalar exchange, because $\omega_3$ is different 
from any prescribed laser energies, {\em i.e.}, neither $\omega$ nor $u\omega$.

The new beam, to be called an {\em inducing} beam, is chosen to share the same 
average number $N$ as before.  This beam  provides us with a sea of photons 
from which the photon $p_4$ is created in the induced manner, 
so that $\bra p_4 |F^{\mu\nu} | 0\ket$ in the second of
\reflef{6d_2}) will be modified to
\beq
\bra p_4 |a^\dagger |0\ket \rightarrow \dbra p_4,N |a^\dagger   |
p_4,N \dket =\sqrt{N},
\label{6d_4}
\eeq
precisely the time-reversed process of the ones for the incident photons
with $p_1$ and $p_2$   in $V_1$, as in \reflef{enhrev1_3}).  Notice that
the photon $p_3$ remains to be created spontaneously from the vacuum because 
its momentum is quite away from $p_4$ of the beam.

No attempt is made to observe the photon $p_4$, which will be embedded
quietly in $\dbra p_4,N|$.  We nevertheless reach a remarkable consequence;
detecting the photon $p_3$ in the unique observational signature as
emphasized before, but with the rate enhanced by $N$, because $\sqrt{N}$
on RHS of \reflef{6d_4}) multiplies the effective coupling strength $B$
in $V_2$ in the second of \reflef{6d_2}) at the second vertex.
Combining this with the previous enhancement by $N^2$ arising from the
first vertex, we finally find that the net enhancement of the rate of
the signal is $N^3$ which replaces $N^2$ in \reflef{exeq_6}).  This
implies that the exponent 1/2 on the middle of \reflef{exeq_9}) is now
replaced by 1/3.  This entails $\sim 10^{21}$ optical photons $\sim
1{\rm kJ}$ on its RHS  for the required pulse energy, fortunately
achievable within the current laser technology.

\bcent
\bfig[h]
\hspace{10.5em}　　　　　
\includegraphics[width=8.0cm]{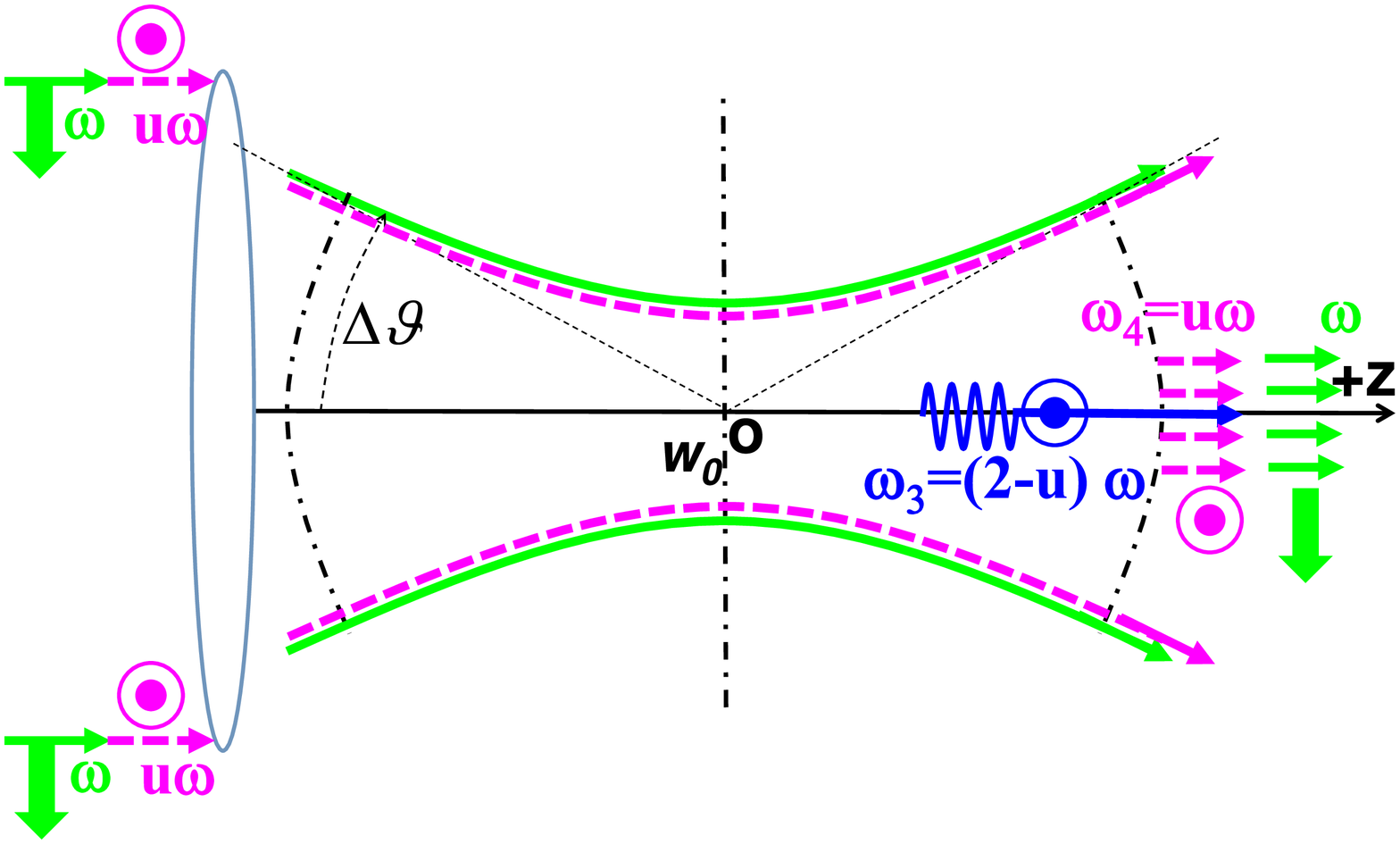}
\mbox{}\\[-2.0em]
\caption{  
A laser field with $\omega$ (solid line) mixed with an {\em inducing} 
laser field with $u\omega$ (dashed line) is focused simultaneously. 
The photon with $\omega_4=u\omega$ is induced to be created from the sea
of photons.  Simply due to the energy-momentum conservation law,
$\omega_3$ is then {\em enforced} to be $(2-u)\omega$ in the forward
direction as shown in Fig. \ref{Fig2}.  We re-emphasize that $p_4$ had
been chosen precisely for this purpose.  We also point out that $p_3$ is
included in none of the momentum ranges of the initial frequency-mixed
lasers. In this way $\omega_3$ remains to be a clear experimental
signature by further requiring the proper polarization state. }
\label{Fig6}
\efig
\ecent 
\mbox{}\\[-3.5em]

We may argue, on the other hand, preparing another incident beam may
provide another type of the 2-photon system with the frequencies $\omega$
and $\omega_4$, coming to the interaction \reflef{mxelm_1}).
Such a two-body system, however, makes an negligible contribution to 
the event rate, if only spontaneous processes are taken into account.
In spite of all such unimportant contributions, we have demonstrated
that adding the inducing beam as prescribed above does yield the
frequency-upshifted photon, as was predicted without the inducing
beam before,  re-emerging simply with the rate enhanced to the level of
physical detectability.

\setcounter{equation}{0}
\section{Conclusion}

We are going to summarize briefly how we have arrived at the final scenario
in which  the gravitationally weak signals are enhanced to the level of
 being subject to measurements, hopefully in the realm of laboratory
 experiments, as will be shown according to the series of steps  below.\\[-.8em]

\noindent\\
(i)\hspace{.8em} We exploited the resonance nature of the scalar field.
We consider the cross section of elastic photon-photon scattering in the
$s$-channel, exhibiting an overall behavior of  ${\cal O}(\MP^{-4})$.
At the resonance precisely corresponding to $m_\phi$, the cross section
reaches  the peak value of   ${\cal O}(\MP^0)$, independent of the
strength of the force.  In this sense we expect an enhancement of the
size ${\cal O}(\MP^4)$, which might be re-expressed by a dimensionless
parameter $(\MP/m_\phi)^4 \sim (10^{36})^4 \sim 10^{144}$.  This
advantage can be offset by the width $a \sim
10^{-77}({\rm eV})^2$, being extremely narrower than any of the
realistic squared-energy scales, chosen to be $\sim ({\rm eV})^2$, for
example.  We then apply certain averaging process over the practical
range, as noted above.  This causes a setback of the enhancement of
$\sim 10^{-77}$, which leaves us nearly a half-way, still above the
bottom by more than 70 orders of magnitude.

\noindent\\
(ii)\hspace{.8em}  Both of the phase-volume integral and the flux of the
photons in the quasi-parallel frame result in the enhancement $\sim
\vartheta^{-2}$, where $\vartheta \sim 10^{-9}$ is for the half the
angle made by the two incident photons.  Out of the expected 
total amount  $\vartheta^{-4}\sim 10^{36}$, a `` half,''
$\vartheta^{-2}\sim 10^{18}$ was shown to be compensated by the small
probability with which we detect the characteristic forward peak of the
nearly frequency-doubled outgoing photon.  The remaining
$\vartheta^{-2}$, however,  does contribute the enhancement by $\sim 10^{18}$.

\noindent\\
(iii)\hspace{.8em} We finally appealed to highly intensified laser beam,
supposed to be described most likely by a coherent state providing with,
 not the vacuum, but a sea of photons which the incident or outgoing
photons are annihilated into or created from.  Due basically to the
induced creation/annihilation mechanism, each of the amplitudes for
three of the external photons is enhanced by $\sqrt{N}$ with $N$ for the
average photon number contained in each of the pulse beam.  The entire
yield ${\cal Y}$ per  pulse focusing for observing the expected 
frequency-shifted photons in the forward direction is then proportional
to $N^3$. We find that ${\cal Y}$ turns out to be of the order
unity if $N$ is as large as $10^{21}$, corresponding fortunately to the
beam supposed to be available soon with the most advanced laser technology.
\\[1.0em]

Summarizing, we point out that we have achieved only what we may call
a simplified overview of the whole proposal.  In the
future studies on the more realistic world, on the other hand, we should
focus upon more details, particularly on how we depart from the
single-frequency mode of the coherent state. This may result in revising the
present estimate for the necessary energy per laser pulse to some
extent, though the  major conclusions reached so far are expected to
remain basically unchanged. \\[1.6em]  

\section*{Acknowledgments}
K. H. thanks D. Habs, R. H\"{o}rline, S. Karsch, T. Tajima,
S. Tokita, L. Veisz and M. Zepf for valuable discussions.
Y. F. is indebted to  Y. Sakurayama, A. Iwamoto, T. Izuyama, H. Matsumoto and
K. Shimizu for their helpful comments.  
This work was supported by the Grant-in-Aid for Scientific
Research no.21654035 from MEXT of Japan in part and the DFG Cluster of
Excellence MAP (Munich-Center for Advanced Photonics).\\[.2em]

\appendix
\renewcommand{\theequation}{\Alph{section}.\arabic{equation}}
\setcounter{equation}{0}
\section{Phase-volume integral}

We start with
\beq
\frac{d\sigma}{d\Omega_3}=\frac{1}{64\pi^2}\frac{1}{\omega^2 \sin^2\vartheta}\int_0^\infty
d\omega_3 \:
\omega_3  \int\frac{d^3p_4}{2\omega_4}\delta^4(p_3+p_4 -p_{\rm
in})|{\cal M}|^2.
\label{app_a2}
\eeq
We insert an identity
\beq
1=\int_0 ^\infty dp_4^0 \delta(p_4^0 -\omega_4)
= \int_{-\infty}^\infty dp_4^0\: 2\omega_4 \delta( p_4^2 )\Theta(p_4^0),
\label{app_a3}
\eeq
where $\omega_4>0$ is always understood in what follows.

In the quasi-parallel frame, we compute
\beqa
 p_4^2 &=&(p_1+p_2 -p_3 )^2=2p_1
p_2-2p_3(p_1+p_2)\nnb\\
&=&4\omega   \left( 1-\cos\vartheta \cos\theta_3 \right)  \left( \omega_3 - \omega \frac{\sin^2 \vartheta}{1-\cos\vartheta \cos\theta_3}  \right),
\label{app_a5}
\eeqa
hence
\beq
\delta\left( p_4^2  \right)= \frac{1}{
4\omega   \left( 1-\cos\vartheta \cos\theta_3 \right) } \delta\left(
\omega_3 - \hat{\omega}_3  \right),
\label{app_a6}
\eeq
where
\beq
\frac{\hat{\omega}_3}{\omega}\equiv \frac{\sin^2\vartheta}{1-\cos\vartheta
\cos\theta_3}.
\label{app_a7}
\eeq
By re-expressing this into
\beq
\frac{1}{1-\cos\vartheta \cos\theta_3}\approx \vartheta^{-2}\frac{\hat{\omega}_3}{\omega},
\label{app_a8}
\eeq
which is substituted into \reflef{app_a6}), further into
\reflef{app_a2}) through \reflef{app_a3}), we finally obtain
\beq
\frac{d\sigma}{d\Omega_3}=\frac{1}{64\pi^2\omega^2}\vartheta^{-2}\int
d\omega_3 \omega_3 \frac{\hat{\omega}_3  }{4\omega^2}
\vartheta^{-2}\delta\left( 
\omega_3 - \hat{\omega}_3  \right)|{\cal M}|^2
=\frac{1}{64\pi^2\omega^2}\vartheta^{-4}\left(\frac{\hat{\omega}_3}{2\omega}\right)^2|{\cal M}|^2,
\label{app_a9}
\eeq
where $\omega_3$, if contained in $|{\cal M}|^2$, is understood to be
$\hat{\omega}_3$, though  such $\omega_3$-dependence rarely occurs in
practice.

We also point out that the invariant amplitude is supposed to represent the
effect of the $\phi$ resonance.  Particularly used in conjunction with
the averaging process explained in section 3, $|{\cal M}|^2$ is
nonzero only for the specific choices of the initial frequency or the
incident angle favoring the resonance condition.  In this sense the
purely kinematical contribution represented by the terms other than
$|{\cal M}|^2$ in \reflef{app_a9}) will be important.

\section{An estimate of the coefficient $\eta$}
\setcounter{equation}{0}

By substituting $\vartheta_\pm$ in LHS of \reflef{exeq_2ccc}), we obtain
\beq
\pm\tilde{a} =\chi(\vartheta_\pm)= \omega_1^2 K_\pm,
\label{app_b1}
\eeq
\beq
 K_\pm = 1- \left(\frac{\vartheta_r}{\vartheta_\pm} \right)^2,
\label{app_b12}
\eeq
By identifying LHS of \reflef{app_b1}) with  $\pm a\eta$, we also obtain
\beq
\eta = \pm\eta_0 K_\pm,
\label{app_b2}
\eeq
where $\eta_0$ is already defined by \reflef{exeq_63_1}), also $\kappa
\equiv B^2/(4\pi) \sim 10^{-5}$.  Notice that  \reflef{app_b2}) consists
of the two equations including the equality 
\beq
K_+=- K_-,
\label{app_b3}
\eeq
a manifestation of the assumed symmetry for simplicity, the same
distances of $\chi$ from the resonance at $\chi =0$ in the integration
range in  \reflef{mxelm_14}).  Using this in \reflef{app_b2}), produces
\beq
\eta =\eta_0\half \left( K_+ - K_- \right)=\eta_0 \half\left(
\left( \frac{\vartheta_r}{\vartheta_-} \right)^2 - \left(
\frac{\vartheta_r}{\vartheta_+} \right)^2 \right). 
\label{app_b4}
\eeq

We also substitute \reflef{app_b12}) into \reflef{app_b3}), obtaining 
\beq
\left( \frac{\vartheta_r}{\vartheta_+} \right)^2+\left( \frac{\vartheta_r}{\vartheta_-} \right)^2=2.
\label{app_b5}
\eeq
By eliminating $\vartheta_-$ from \reflef{app_b4}) and \reflef{app_b5}),
we finally obtain
\beq
\eta =\left( 1- \left( \frac{\vartheta_r}{\vartheta_+}
\right)^2\right)\eta_0, 
\label{app_b6}
\eeq
hence deriving \reflef{exeq_63}).  

\section{Averaging over $\vartheta$}

\setcounter{equation}{0}

We consider
\beq
\overline{|{\cal M}_{1111s}|_\vartheta^2}=\frac{{\cal N}_\vartheta}{{\cal D}_\vartheta},
\label{app_bc1}
\eeq
where
\beq
{\cal N}_\vartheta =\int_{\vartheta_-}^{\vartheta_+}|{\cal M}_{1111s}|^2
d\vartheta,\quad\mbox{and}\quad {\cal D}_\vartheta = \vartheta_+-\vartheta_-.
\label{app_bc2}
\eeq

On RHS of the first of \reflef{app_bc2}), we substitute the Jacobian $J$
to obtain
\beq
d\vartheta =Jd\chi,
\label{app_bc3}
\eeq
where
\beq
J =\left(\frac{d\vartheta}{d\chi}\right) =\left( \frac{d\chi}{d\vartheta} \right)^{-1}=\frac{\vartheta^3}{2\omega_1^2\vartheta_r^2},
\label{app_bc4}
\eeq
as has been obtained from \reflef{exeq_2ccc}).  We substitute
\reflef{app_bc3}) into \reflef{mxelm_14}), in which the integrand is
dominated by the value at $\chi =0$, corresponding to $\vartheta
=\vartheta_r$ according to \reflef{exeq_2ccc}).  We then find that RHS
of  ${\cal N}_\vartheta$ in the first of \reflef{app_bc2}) is
$\vartheta_r/(2\omega_1^2)$, the value of  \reflef{app_bc4}) at $\vartheta
=\vartheta_r$, times the value on RHS of $2\tilde{a}\overline{|{\cal
M}_{1111s}|^2} $.  In this way we obtain
\beq
\frac{\overline{|{\cal M}_{1111s}|_\vartheta^2}}{\overline{|{\cal
M}_{1111s}|^2}}\approx \frac{1}{{\cal D}_\vartheta}\frac{\vartheta_r}{2\omega_1^2}2\tilde{a} \approx \frac{\vartheta_r}{\vartheta_+},
\label{app_bc5}
\eeq
where we have used ${\cal D}\approx \vartheta_+$, as well as
\beq
\tilde{a}=a\eta \approx a\eta_0 =\omega_1^2,
\label{app_bc6}
\eeq
in accordance with \reflef{exeq_63_1}).

\section{Partially integrated cross section, $\bar{\sigma}\sim
 \vartheta^{-2}$ }

\setcounter{equation}{0}

From \reflef{kinm_13}) and \reflef{kinm_14}), we derive
\beq
\int_0^{\bar{\theta}_3}\left( \frac{\bar{\omega}_3}{2\omega}
\right)^2\sin\theta_3 d\theta_3 =\frac{\vartheta^4}{4}\int_{\underline{x}}^1
\left( 1-x\cos\vartheta\right)^{-2}dx \equiv\frac{\vartheta^4}{4}
\int_{\underline{x}}^1 f(x)dx,
\label{app_c1}
\eeq
where $f(x)$ has its indefinite integral, such that $f=d{\cal F}/dx$
with $x=\cos\theta_3$;
\beq
{\cal F}(x)=\frac{1}{\cos\vartheta}\left( 1-x\cos\vartheta \right)^{-1}\approx \left( 1-x\cos\vartheta \right)^{-1},
\label{app_c2}
\eeq
with $\cos\vartheta \sim 1$ in the denominator since the correction term
$\sim \vartheta^2$ has been already in \reflef {app_c1}) as a
multiplicative factor.  We readily find
\beq
{\rm \reflef{app_c1})} =\frac{\vartheta^4}{4}\left( {\cal F}(1)-{\cal F}(\underline{x})
\right)\approx  \frac{\vartheta^4}{4} \vartheta^{-2}\left(
2-\frac{\bar{\omega}_3}{\omega}\right)=\frac{1}{4}\vartheta^2 u,
\label{app_c3}
\eeq
thus yielding \reflef{angav_16}), where \reflef{app_c2}),
\reflef{app_a8}) with $\theta_3, \hat{\omega}_3$
replaced by $\bar{\theta}_3, \bar{\omega}_3$, respectively,  and
\reflef{angav_14}) have been used.  This provides us with
\reflef{angav_16}).

For a certain experimental condition, we may need the
integral in \reflef{app_c1}), but with the lower bound 0 replaced by
$\underline{\theta}_3$.  In order to estimate ${\cal F}(x)$ in a more general
value of $x$, we find it convenient to introduce $v_3$ by
\beq
\theta_3=v_3\vartheta, \quad\mbox{hence}\quad u=2v_3^2.
\label{app_c4}
\eeq

We then obtain
\beqa
{\cal F}(x)&\approx & \left( 1-x\cos\vartheta \right)^{-1} =\left( 1-\left(
1-\half v_3^2\vartheta^2 \right)\cos\vartheta  \right)^{-1} \nnb\\
&=&\left( 1-\cos\vartheta +  \half v_3^2\vartheta^2 \cos\vartheta
\right)^{-1} \approx \left( \half \vartheta^2 +  \half v_3^2\vartheta^2
\cos\vartheta  \right)^{-1} \nnb\\
&=& \left( \half \vartheta^2 (1+v_3^2)
\right)^{-1}=2\vartheta^{-2}\frac{1}{1+v_3^2} \approx
2\vartheta^{-2}\left( 1-v_3^2\right).
\label{app_c5}
\eeqa
According to \reflef{app_c1}) we find
\beq
\bar{\sigma}=\frac{\pi^2}{16\omega^2}\eta^{-1}\left( {\cal
F}(\underline{x})-{\cal F}(\overline{x}) \right) \approx
\frac{\pi^2}{16\omega^2}\eta^{-1}2\vartheta^{-2}\left( \overline{v}_3^2 - \underline{v}_3^2  \right).
\label{indb_c6}
\eeq


　　　
\section{Understanding the enhanced yield in the quasi-parallel frame}
\setcounter{equation}{0}

The cross section $\bar{\sigma}$ is very large due to the factor $\sim
\vartheta^{-2}\sim 10^{18}$ in \reflef{angav_17}), making a significant
contribution to the 
value of the final yield ${\cal Y}$ according to \reflef{coh_21}).  We
should make sure, however, that the calculation correctly ensures that
we avoid to fall into the same path as another example of $\vartheta^{-2}$
which was shown to be compensated away by a small partially-integrated
cross section discussed toward the end of section 4.  For this
purpose, it might be helpful if we better understand how the relation
\reflef{coh_21}) is derived applied particularly to scattering of
coherent states.  The argument will be made most conveniently by
identifying the 
enhancement $\sim \vartheta^{-2}$ under discussion to come from the flux
in computing the elastic photon-photon scattering cross section in the
quasi-parallel incident frame.

We start with explaining how we compute the flux in the photon-photon
scattering cross section based on Mo\hspace{-.5em}/ller's formulation
which may be found in (8-50) of \cite{JR}, or (3.78) of \cite{GR}.

Define a Lorentz-invariant function
\beq
F=\sqrt{\left( p_1p_2 \right)^2 -m^4},
\label{app_e1}
\eeq       
for two identical massive particles of the momenta $p_1$ and $p_2$.  We
may demonstrate that \reflef{app_e1}) is justified in the number of
well-known examples.

First in the center-of-mass frame with $\vec{p}_1=-\vec{p}_2
=\vec{p},\quad p_1^0 =p_2^0 =\omega$, we derive easily $F=\omega^2
v_{\rm rel}$, where $\vec{v}_{\rm rel}=\vec{v}_1 -\vec{v}_2 = 2\vec{p}/\omega$, with $\vec{v}_i=\vec{p}_i/\omega$ (i=1,2).

Likewise, in the laboratory frame we have $\vec{p}_1 =\vec{p},
\omega_1=\omega$, and $\vec{p}_2 =0, \omega_2=m$,  from which follows $F=m\omega_1 v_1 =m\omega_1 v_{\rm rel}$.

These exercises indicate a computational rule that the familiar normalization factor $(\omega_1 \omega_2)^{-1}$ also divided by $v_{\rm
rel}$ should be replaced by a Lorentz-invariant factor $F^{-1}$.

Now we go to the massless limit $m\rightarrow 0$ in \reflef{app_e1}).  In
the ``center-of-mass'' frame with $\vec{p}_2 =-\vec{p}_1,\quad
\omega_2=\omega_1,$ we compute
\beq
|p_1p_2|=|-\vec{p}_1^{\;2}-\omega_1^2|=2\omega^2 =\omega^2\times (1+1),
\label{app_d7}
\eeq
where we interpret $(1+1=2)$ reasonably as the relative velocity of the
massless particles.  After these preparations, we finally consider the
quasi-parallel geometry as was discussed in section 2;
\beq
\vec{p}_{2x}=-\vec{p}_{1x},\quad \vec{p}_{2z}=\vec{p}_{1z},\quad \omega_2=\omega_1,
\label{fl_12}
\eeq
from which follows immediately 
\beq
F=|p_1p_2|=|-\omega^2 \sin^2\vartheta +\omega^2 \cos^2\vartheta
-\omega^2| =\omega^2 |-2\sin^2\vartheta| =\omega^2 2\sin^2\vartheta,
\label{fl_13}
\eeq
establishing what should be re-interpreted as the relative velocity in
the present frame is given by $2\sin^2\vartheta \sim 2 \vartheta^2$.

Usually the cross section obtained for a unit flux is multiplied with
another flux to be determined depending on the physical
circumstance.  In the current beam-type experiment,  as illustrated in
Fig. \ref{onebeam}, the photons are supplied from the left in the
positive direction of the horizontal $z$-axis.  This remains true even for two
incident photons which will be bent afterward by the lens.  In this
sense we have no relative velocity prepared at the observational
entrance doorway.

We then naturally adopt the equivalent concept of a ``flow,'' which
shares the same property with a flux; the number of particles
passing through unit perpendicular area per unit time, but without
relying on the relative velocity.  The area perpendicular to the flow is
chosen to be $\pi w_0^2$, where $w_0$  is the waist of the laser beam.  
We notice a decisive difference from the way we defined the flux in the
microscopic range in which the relative velocity is obviously
perpendicular to the $z$-axis.  We find no way for the two ways to
affect each other.

A crucial observation is that by particles mentioned above we never mean the
individual photons which are uncountable, but should understand to be
the coherent states described by \reflef{coh_2}).  The flow thus represents
the number of these coherent states per perpendicular area per unit
time.  We prepared the initial state as a single number of the coherent
state, with the response to the scalar field interaction basically given
by  \reflef{coh_3d}).

We also notice that we are going to estimate the yield per pulse
focusing, instead of the rate per unit time.  We are integrating
the flow over the entire time-span of the pulse, giving the number unity
for the coherent state.  This is the way we obtained $I_c$ by a product
of unity times $N^2$ in \reflef{coh_22}).

After all we are left with none of the variables responsible for any
 of the  time-dependence, like the flow, leading to the simple result
\reflef{coh_21}) with \reflef{coh_22}), in which the large value
coming from $\vartheta^{-2}$ in $\bar{\sigma}$ is robust.


 \end{document}